# Exploring the Impact of Generative Artificial Intelligence on Software Development in the IT Sector: Preliminary Findings on Productivity, Efficiency and Job Security

*Anton Ludwig Bonin*
*University of Gdańsk, Faculty of Economics*
*Gdańsk, Poland*                    *l.bonin.707@studms.ug.edu.pl*

*Pawel Robert Smolinski*
*University of Gdańsk, Faculty of Economics*
*Gdańsk, Poland*                    *pawel.smolinski@ug.edu.pl*

*Jacek Winiarski*
*University of Gdańsk, Faculty of Economics*
*Gdańsk, Poland*                    *jacek.winiarski@ug.edu.pl*

**Abstract**

This study investigates the impact of Generative AI on software development within the IT sector through a mixed-method approach, utilizing a survey developed based on expert interviews. The preliminary results of an ongoing survey offer early insights into how Generative AI reshapes personal productivity, organizational efficiency, adoption, business strategy and job insecurity. The findings reveal that 97% of IT workers use Generative AI tools, mainly ChatGPT. Participants report significant personal productivity gain and perceive organizational efficiency improvements that correlate positively with Generative AI adoption by their organizations ($r = .470$, $p < .05$). However, increased organizational adoption of AI strongly correlates with heightened employee job security concerns ($r = .549$, $p < .001$). Key adoption challenges include inaccurate outputs (64.2%), regulatory compliance issues (58.2%) and ethical concerns (52.2%). This research offers early empirical insights into Generative AI's economic and organizational implications.

## 1. Introduction

Technology reshapes economic environments, market dynamics, and employment relationships as it continues to spread across industries [7]. A key feature of this change is the emergence of technologies such as Artificial Intelligence (AI), which can drive improvements in efficiency, automate work and offer new means of creating value [9]. These dynamics are what digital economics aims to study, particularly in terms of how digital technologies, applications, and solutions affect individual and collective participation and exchanges within the struggle for resources. In this context, Generative AI (GenAI) has emerged as one of the most popular fields of study due to its flexible applications. The report *State of Generative AI* [11] from Deloitte and other studies found that organizations are leveraging GenAI for diverse applications, from IT and cybersecurity to code generation, writing, copywriting, marketing content creation [15], [30], customer service and R&D, demonstrating its adaptability across core business functions.

Generative AI can be defined as a type of Artificial Intelligence whose models are optimized for content generation based on data provision. Unlike other forms of AI that are usually boxed into certain categories of applications or problem-solving approaches, Generative AI can create outputs in various formats. These capabilities are valuable in the digital economy because they offer previously unknown approaches for automating creative processes, as well as methods for creating content and software at high speed and, in some cases, stimulating creativity in industries that have traditionally relied on the input of experts [13]. As Generative AI tools gain global adoption, it is important to determine the impact of these tools on roles and responsibilities, productivity, and competitiveness of the particular sectors [16], [32]. Studies like Rajbhoj et al. [26] and Bubeck et al. [6] show how GenAI automates code generation, yet Russo [27] warns about the organizational complexity of integrating these tools. This study focuses on discussing Generative AI in the context of the software development field. There are many methodologies to develop software such as waterfall or agile methods. In reality, regardless of the development methodology chosen, the process involves complex activities and an estimated development budget can be set based on extensive personnel time and technical know-how [31]. To the best of our knowledge, the utilization of Generative AI techniques across various stages of the Software Development Life Cycle (SDLC) remains a relatively unexplored area [26]. There are already studies focusing on particular aspects of SDLC, for example by Russo focusing on adoption issues of Generative AI [27].

Zhang [33] and Li [18] outline how GenAI enhances firm-level productivity, Czarnitzki et al. [9] show that realizing productivity gains from AI adoption depends on firms' organizational readiness and integration efforts, whereas Pothukuchi et al. [25] caution that the economic returns depend heavily on organizational readiness and process integration. Recently, new studies regarding economic effects have been published, such as Zheng [34]. This study mentions that Generative AI reshapes labor markets, boosts productivity and drives economic growth but only if long-term benefits are managed effectively. This also demonstrates the importance of focusing on one narrow sector instead of a broad analysis to investigate challenges, strategies, labor market changes by personal perception and, to establish the circumstances in the sector.

This study focuses specifically on the IT sector to ensure relevant and targeted findings. We selected this sector because Generative AI is expected to have one of its most significant impacts on software development and related IT activities [28]. Although prior work examines technical capabilities [1], [6] and sector applications [25], [28], empirical analysis of GenAI's economic consequences in software development remains limited [8], [32]. There may also be differences of software development in various sectors such as healthcare or supply chain due to preexisting tools, as Zheng demonstrates the importance of the preexisting tools [34].

With this in mind we conducted a pilot study examining the transformative impact of Generative AI on software development in the IT sector. We investigate relationships between GenAI adoption, organizational efficiency, personal productivity, and job insecurity. We aim to quantify the connections between organizational GenAI implementation, productivity improvements, and job market impact. Through this investigation, we seek to establish how Generative AI revolutionizes software development within the IT sector. We also analyze the key barriers preventing full adoption, including output accuracy issues, regulatory compliance challenges, and ethical considerations.

## 2. Generative Artificial Intelligence in Software Development

Generative AI reshapes software development in many ways. In this section, we categorize these changes into economic aspects and market shifts, followed by a discussion of research gaps.

### 2.1. Economic Aspects

Generative AI models are suitable for use in various fields; however, their influence is highly significant in software development. Fortunately, Generative AI has the potential to alter this dynamic by capturing certain elements of software design, like coding and debugging [6]. One of the biggest economic shifts due to Generative AI in software development is the change in productivity [8]. Productivity improvements from GenAI are widely discussed in the literature [1], [8], [33], though their measurement and organizational interpretation remain unclear [28]. This shift is mainly driven by Generative AI's capability to scan large codes, recognize patterns and then generate new code independently. This, in turn, makes the software development cycle quicker, especially writing code, debugging or improving algorithms — tasks that previously required human effort but are now proficiently handled by Generative AI tools like OpenAI's ChatGPT and Anthropic's Claude [1], [26]. For instance, tasks that previously took several days of repetitive manual coding can now be completed within hours using Generative AI tools. This is a huge advantage for organizations since time spent on development can be cut, and expenditures on new products can be reduced, making it easier for businesses to meet the requirements of consumer demand more effectively. Generative AI can rewrite the way applications are written, updated, and possibly scaled through its capability for code sourcing, proposing optimizations and even developing new complete modules of software. Generative AI frees up the employees' time by automating manual tasks. This can result in increased personal productivity, organizational efficiency, improved perceptions of innovation, and can even enable organizations to offer new products and services. Research agrees that GenAI boosts coding efficiency [1], [8], [26], but there is less consensus on whether this translates into sustained innovation or merely short-term productivity gains. This alone can potentially increase the revenue of organizations using Generative AI.

**2.2. Market Shifts**

Software development has depended to a very large extent on human resources for writing, finding errors and fixing programs. However, current trends associated with Generative AI, particularly its application in enhancing and automating different phases of the SDLC, are somewhat ambiguous and impact the demand for skilled software developers across various areas of the IT industry [25], [27]. On the one hand, there is demand for developers who can properly utilize AI and integrate it into their workflows, as companies seek individuals with expertise in both software development and AI capabilities [2]. However, Generative AI provides automation capabilities that may gradually reduce developers' need to perform repetitive coding tasks, potentially changing the types of jobs available. Khan et al. [16] and Piton [24] warn about job displacement, in contrast, Alekseeva et al. [2] and Zheng [34] stress the emergence of hybrid roles blending programming with AI tool management. Of all the changes in skill demands, the most significant is the shift toward working across multiple subject matters. As Generative AI programs become more popular, it is becoming equally critical for software developers to know how to apply Generative AI in their work as well as being good programmers themselves. GenAI holds promise for increased productivity, but there are valid concerns about worker displacement in the IT sector [16]. This transformative effect of Generative AI resembles Schumpeter's concept of "creative destruction" highlighting how innovation simultaneously dismantles established processes while creating new opportunities [29].

**2.3. Research Questions**

Generative AI has become an essential topic in software economics; however, several knowledge gaps remain. The primary issue is the lack of academic studies examining how Generative AI affects the economic architecture of the SDLC. Since Generative AI research is relatively recent, most studies focus on technical aspects and general industry applications rather than economic consequences such as productivity improvements, organizational efficiencies, and innovation strategies within software development [1], [32]. There is a strong need for research combining quantitative productivity insights with qualitative perspectives from practitioners [8], [24], [32].

The impact of Generative AI on job dynamics—including the balance between role displacement and creation—has not been fully explored [24], [32]. Significant research deficiencies exist regarding labor market trends in the IT sector influenced by Generative AI. As developer requirements change and new roles emerge to work with AI-generated code and design solutions, there is a corresponding shift in skill demands. This highlights the need for research examining how developers adapt to new workflows and how organizations can effectively manage transitions while maximizing economic benefits. Current research either narrowly focuses on demand for AI-related skills [2] or broadly addresses macroeconomic consequences of workforce automation [16], [34], leaving a gap in between.

Limited empirical studies examine real-world applications of Generative AI in software development, creating a gap in understanding the technology's direct impact on business operations. While case studies of AI-driven firms exist, systematic research quantifying productivity gains, cost reductions, and innovation outcomes attributable to Generative AI remains sparse. This gap is particularly evident in the absence of studies integrating quantitative data with qualitative insights from industry practitioners. To address the identified research gaps, we investigate six research questions:

- **RQ1.** To what extent and in what forms has Generative AI been adopted in software development?
- **RQ2.** Which Generative AI use cases and adoption patterns predominate?
- **RQ3.** How does Generative AI adoption affect individual developer productivity?
- **RQ4.** What is the relationship between Generative AI adoption and organizational efficiency?
- **RQ5**. How does Generative AI adoption influence employee perceptions of job security?
- **RQ6**. What challenges and barriers impede Generative AI integration?

## 3. Methodology

One of the biggest challenges in this research was defining important and researchable topics in the field of GenAI. To define these topics, set research goals, and conduct research, we used mixed research methods to combine the strengths of both qualitative and quantitative approaches. We aimed to create themes without any pre-existing concepts but later test these developed themes with a broader sample. Our mixed research approach begins with a preliminary qualitative analysis followed by a quantitative analysis. This methodology is described in Morgan's four quadrant typology [20] and is also recommended as a method for developing a quantitative survey [14]. This methodology was further extended by Morse [21] through theoretical orientation and timing. In this study, we used an inductive sequential mixed model.

An inductive research approach was essential for exploring unknown processes, generating new hypotheses, and capturing new perspectives without initial conceptions [22], thus matching the exact requirements of this study. To start our inductive qualitative research, we selected the thematic analysis method [5] because it fundamentally aligns with inductive research since it allows creating hypotheses without needing pre-existing concepts and offers considerable flexibility. To collect the qualitative data, three expert interviews were conducted in a semi-structured way, as semi-structured interviews offer the flexibility described by Patton [22] and also by Braun and Clark [4], who established the fundamentals for thematic analysis concepts. We conducted interviews with two senior technical specialists and one software development manager in a combination of online and face-to-face sessions, audio recorded and verbatim transcribed. All transcripts were thoroughly anonymized by removing personal and organizational identifiers to meet interviewees' organizations' confidentiality needs before being imported into MAXQDA for the six-phase thematic analysis. After collecting the qualitative data, we performed the six steps of the thematic analysis as illustrated in Fig. 1.

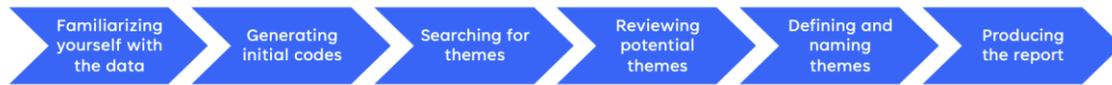

**Fig. 1.** Six-phase thematic analytic process [4, 5]

After successfully conducting and analyzing the expert interviews, we generated numerous potential themes. We then conducted an extensive literature review based on these potential themes. As a result, we selected scientifically interesting and impactful hypotheses as the focus of this research, which coalesced around five key areas: personally perceived productivity, organizational perceived efficiency, organizational adaptation, business strategy and job market impacts.

Based on the defined research questions, we developed a quantitative survey following the survey research guidelines provided by Krosnick and Presser [17], and Brace [3]. The survey uses a Likert scale [19] to understand participants' agreement levels with key arguments. A 5-point Likert scale is a proven method to measure attitudes with equal intervals and provides a balanced scale between reliability and simplicity, as demonstrated by Dawes [10].

The survey begins with two screening questions followed by eight content sections. The screening questions verify whether participants work in the IT sector and whether they participate in an organization that develops software. A negative response to either question terminates the survey. The survey consists of multiple Likert-scaled questions within each section: personally perceived productivity, organizational perceived efficiency, organizational adaptation, business strategy, job market impacts and personal satisfaction, followed by demographic questions. Fig. 2 presents the final outline of the designed survey.

From the questions in these sections, we created five composite scores that capture key dimensions of GenAI's impact in organizational settings: Organizational adoption, perceived productivity, organizational efficiency, business strategy and job insecurity. The organizational adoption score measures how extensively organizations have implemented Generative AI solutions. Perceived productivity assesses respondents' perceptions of how AI affects their work efficiency and quality, with higher scores indicating more positive evaluation. Organizational efficiency evaluates perceived impacts on organizational performance, costs and innovation capacity. Higher scores indicate better improvements in the organization due to Generative AI. The business strategy score assesses the degree of resource allocation for AI initiatives. The higher the score, the more strategic prioritization of Generative AI initiatives within the organization. It is a similar score to organizational adoption; however, organizational adoption measures more current implementation of Generative AI, whereas the business strategy score measures willingness to invest and prioritize Generative AI initiatives in the future. Finally, the job insecurity score incorporates items to measure perceptions about how Generative AI affects employment opportunities and skill requirements. Higher scores on the job insecurity scale indicate perception of more negative impact of AI on one's job security. In this study, all scales demonstrated acceptable validity and reliability. Table 1 presents the McDonald's omega and Cronbach's alpha for all the scales used in the study.

**Table 1.** McDonald's omega and Cronbach's alpha for all the scales used in the study

|  | McDonald's $\omega$ | Cronbach's $\alpha$ |
| --- | --- | --- |
| Productivity | .860 | .846 |
| Organizational Efficiency | .663 | .609 |
| Organizational Adaptation | .745 | .725 |
| Business Strategies | .938 | .937 |
| Job Insecurity | .662 | .605 |

We conducted this initial survey as a pilot with 12 participants who provided direct feedback on survey design, including identifying unclear questions, options and offering other suggestions. Based on this feedback, we clearly defined the sections and adjusted the questions and options. After implementing these changes, we finalized the survey. To gather more generalizable results, we conducted the finalized survey in one country, Poland, and made it available in two languages: Polish and English. The data we present in the following section represent preliminary results obtained from this survey. We are preparing a follow-up publication that will include full methodological exposition. As the instrument is still being used in ongoing data collection, we prefer to reserve complete methodological disclosure for that forthcoming paper.

## 4. Results

This section presents the preliminary findings of our survey on Generative AI adoption in the IT sector. The survey yielded 96 completed questionnaires. We removed respondents who did not meet our screening criteria of being employed in the IT sector and working in software development roles. In total 28 respondents were filtered out, leaving 68 qualified cases for analysis. The final sample is largely male (94%), highly educated (~60% hold a master's degree and 25% hold an engineer's degree) and concentrated in the 25–44 year age brackets (81%). The participants come from organizations of every size (38% micro, 15% small, 28% medium, 19% large) and over a third (38%) describe their organizations as digitally advanced, with a further 28% as digitally maturing.

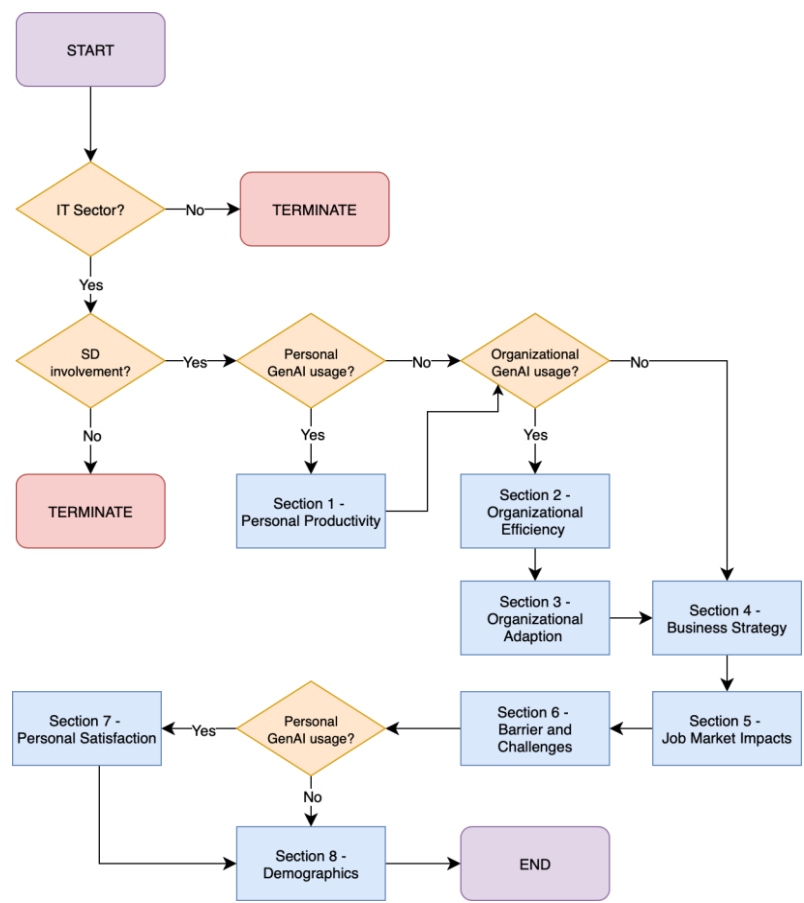

**Fig. 2.** The process flowchart of the designed survey instrument

## 4.1. Generative AI Models Popularity

97% of surveyed respondents (66 respondents) admit to using Generative AI tools to some extent for their job. Fig. 3 presents the popularity of each Generative AI tool among surveyed respondents. ChatGPT with its series of models is the most popular, likely due to its pioneer status as the first to reach wide public adoption. GitHub Copilot, DeepSeek, and Google's Gemini models follow closely in popularity. Copilot's popularity may result from its seamless integration with code development software, whereas Gemini's popularity may stem from its integration with Google products often bundled with AI. Interestingly, other high-performing but closed-source models like the Claude series (Anthropic) and the Grok series (xAI models) are relatively less popular. LLaMA and custom models are the least popular among the surveyed respondents, possibly due to LLaMA's lack of a custom web application and restricted use in the EU.

The percentages in Fig. 3 exceed 100%, indicating that often more than one model is used by respondents. This multi-product usage has likely contributed to the rise of third-party providers like Mammoth AI or T3Chat, which offer Generative AI models from different providers in one service. These platforms are a convenient solution for users who want to access multiple models without the hassle of managing separate accounts. However, we did not survey our participants for the usage of these third-party solutions.

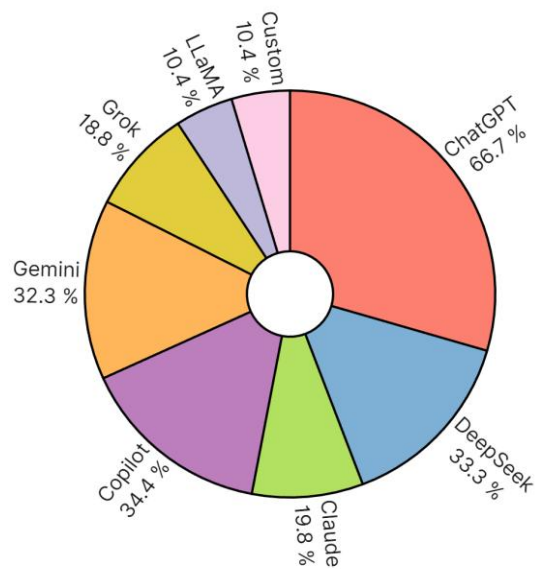

**Fig. 3.** Distribution of Generative AI usage frequency

## 4.2. Generative AI Usage Purpose

Figure 4 presents a bar chart with the percentage of respondents using Generative AI models for various tasks. Copywriting and learning are the most common uses of Generative AI among surveyed participants. Copywriting encompasses categories like writing emails, reports, presentations, and other duties requiring written content creation. There are no differences for these categories between software engineer/developer respondents and manager/executive respondents. For both groups, copywriting and learning are major applications of Generative AI in their jobs. Code generation is the third most reported application, with significant differences between engineers/developers and managers/executives, with developers using it significantly more often ($\chi^2 = 4.267$, p = .039). Similar significant differences are observed for code optimization ($\chi^2 = 4.889$, p = .027), code review ($\chi^2 = 9.771$, p = .002), and code testing and quality assurance ($\chi^2 = 5.664$, p = .017). Code debugging is in the lower half of usage percentage, and surprisingly, there are no significant differences between managers and engineers/developers in their likelihood of using Generative AI for this task. These findings suggest that software engineers and developers are more likely to utilize Generative AI to automate the initial code generation process and then manually debug the code themselves, rather than writing the code manually and asking Generative AI to debug it.

Beyond code-related tasks, managers and executives are finding value in applying Generative AI to project management ($\chi^2 = 6.776$, p = .009) and internal technical drafting ($\chi^2 = 4.080$, p = .043) compared to engineers and developers. In summary, copywriting and learning are the most common applications of Generative AI models for both software engineers/developers and managers/executives. There are significant differences in their usage patterns for code-related tasks and project management. For all other tasks not specifically mentioned, there are no significant differences between the two groups.

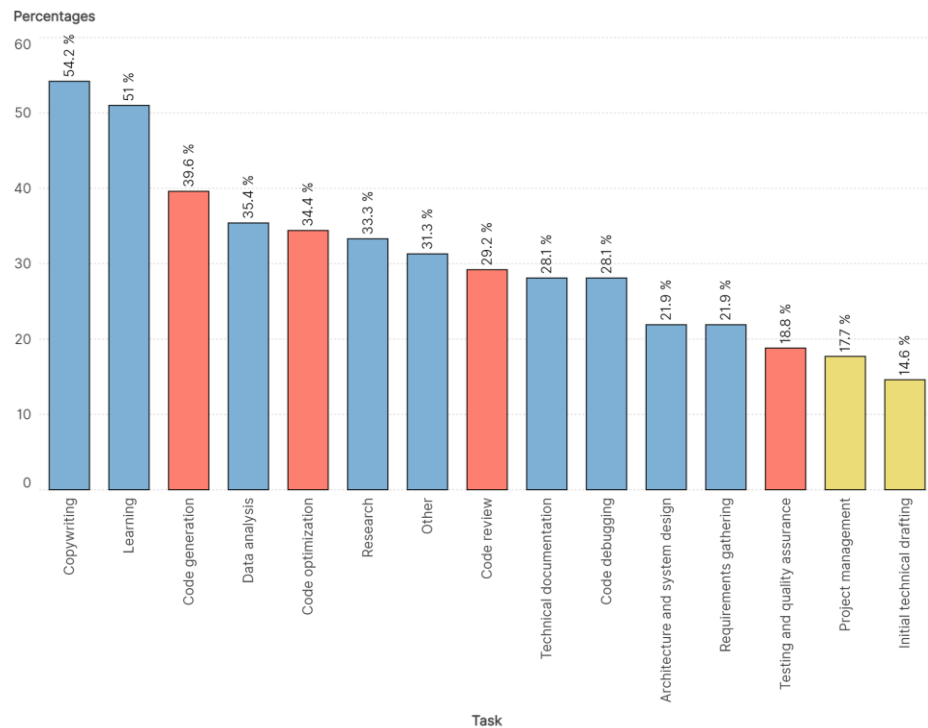

**Fig. 4.** Percentage of respondents using Generative AI models for various tasks by user group (blue = equal; red = higher in technical jobs; yellow = higher in managerial jobs)

### 4.3. Impact of Generative AI on Productivity, Organizational Efficiency and Job Insecurity

Table 2 presents Spearman's correlations between the composite scores. Productivity showed a moderate positive correlation with organizational efficiency (r = .448, p < .01) and organizational adoption (r = .395, p < .05). Organizational adoption also correlated with organizational efficiency (r = .470, p < .01). Business strategy had a strong positive correlation with job insecurity (r = .549, p < .001). Additionally, job insecurity showed a moderate positive correlation with organizational adaptation (r = .403, p < .05).

**Table 2.** Correlations among composite scores of Generative AI's impact on productivity, organizational efficiency and adaptation, business strategy and job insecurity

|  | Productivity | Organizational Efficiency | Organizational Adaptation | Business Strategy |
|---|---|---|---|---|
| Productivity | — | | | |
| Organizational Efficiency | .448** | — | | |
| Organizational Adaptation | .395* | .470** | — | |
| Business Strategy | .230 | .412* | .205 | — |
| Job Insecurity | .236 | .288 | .403* | .549*** |

Note: * $p < .05$, ** $p < .01$, *** $p < .001$

These findings suggest that organizations with higher levels of Generative AI adoption and implementation tend to perceive greater improvements in efficiency and productivity. However, the same organizations that prioritize and invest in AI initiatives are more likely to anticipate negative impacts on employment opportunities and job security. The strong correlation between business strategy and job insecurity scores indicates that as organizations allocate more resources to AI initiatives, concerns about job displacement and changing skill requirements become more prominent. This suggests a possible trade-off between the benefits and drawbacks of AI adoption in organizations. As organizations adopt and benefit from Generative AI solutions, the perceived threat to job security tends to increase among employees. On the other hand, the correlation between Job Insecurity and Productivity is not significant ($r = .236$, ns). This suggests that job insecurity might be present only among employees who do not perceive themselves to be productive with Generative AI tools and are employed in organizations that are adopting AI.

### 4.4. Key Challenges, Barriers, and Risks Associated with the Use of Generative AI

Figure 5 presents the key challenges, barriers, and risks associated with the use of Generative AI models as reported by survey respondents. The top concern, cited by 64.2% of respondents, is inaccurate outputs. This suggests that a majority of IT professionals are worried about the reliability of AI-generated code and content. As Generative AI hallucinations could lead to errors, bugs, or inconsistencies that result in major project setbacks, avoiding them is a priority for many Generative AI users. The second most common concern (58.2%) is regulatory compliance challenges, indicating apprehension about meeting legal and industry standards when incorporating Generative AI into workflows.

Ethical concerns (52.2%), data leakage risk (44.8%) and lack of trust in AI (38.8%) are the next most frequently reported challenges. These findings indicate that organizations recognize the potential of Generative AI, however they are also concerned about responsible AI development, data privacy and security, and the overall trustworthiness of AI systems. Other barriers include insufficient data availability (32.8%), uncertain ROI (28.4%), and resistance to adoption (23.9%). Interestingly, issues like limited automation capabilities (20.9%), skilled staff shortage (20.9%), high operational costs (17.9%), inadequate infrastructure (16.4%) and insufficient AI performance (10.4%) are among the least cited challenges, suggesting that the technical capabilities and resources are not the primary inhibitors to Generative AI adoption in the IT sector.

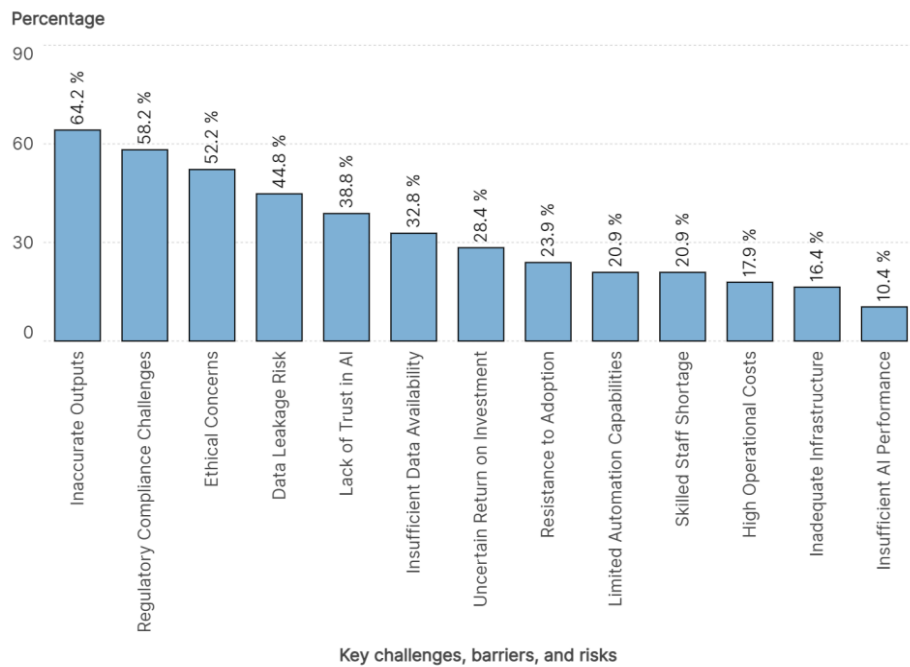

**Fig. 5.** Key challenges, barriers and risks associated with Generative AI usage in software development

## 5. Discussion

This preliminary study, guided by an inductive sequential mixed-method approach, examined how GenAI reshapes software development in the IT sector. The results indicate 97% of respondents use GenAI tools like ChatGPT, Copilot and Gemini and report productivity gains (RQ1). These productivity gains align with the dominant GenAI adoption rate. Copywriting and learning are the top GenAI uses in technical and managerial roles. Engineers use AI significantly more for code generation, optimization, review and testing. Managers use it more for project management and technical drafting. Despite these differences in specific tasks, both groups rely heavily on AI for writing content and learning (RQ2).

　　A significant correlation exists between organizational GenAI adoption rates and perceived improvements in productivity ($r = .395$, $p < .05$) and organizational efficiency ($r = .470$, $p < .01$). This relationship indicates that organizations adopting GenAI more extensively perceive greater benefits in the form of more productive employees (RQ3) and efficient internal workflows (RQ4). A strong positive correlation ($r = .549$, $p < .001$) was found between organizational investment in GenAI and employee concerns about job insecurity, suggesting that increased emphasis on GenAI-driven strategies may lead to greater employee uncertainty regarding their job roles (RQ5). Employees may see GenAI not just as a productivity tool but as a potential substitute for some tasks. Previously a company might have needed many engineers of different qualifications to do both routine and complex tasks, and now a significantly smaller team can accomplish more thanks to the automation of routine tasks with GenAI.

　　We also identified prominent barriers to GenAI adoption, including inaccurate AI outputs (64.2%), regulatory compliance challenges (58.2%), and ethical considerations (52.2%). These barriers significantly influence adoption attitudes and require thorough investigation to develop effective organizational adoption frameworks (RQ6).

　　While the thematic interviews and pilot survey provided useful insights, this study presents limitations and these will be addressed in the future study with more detailed methodology, theoretical background and using finalized survey data. First, our survey is exploratory and captures a specific moment in the evolution of GenAI in the IT sector. Second, the sample size is relatively small (68 valid responses). Third, this pilot relies on self-reported productivity ratings gathered via Likert-scale items, as collecting objective coding metrics such as commit counts and issue-resolution times was outside its scope. However, self-assessed productivity measures have been validated in empirical software engineering research [12], [13]. Furthermore, we acknowledge that some demographic groups may be under-sampled in the current stage of the study. For example, our sample is 6% women, while national IT workforce statistics for Poland according to Eurostat 2024 show that women account for roughly 18% of IT professionals [35]. The data collection is still ongoing to improve generalizability.

## 6. Conclusion

In line with our initial aims, this research offers early empirical evidence that Generative AI can deliver measurable productivity and efficiency improvements in software development, though accompanied by genuine workforce insecurities. Through a mixed-method approach combining expert interviews and survey data, the study reveals near-universal adoption of GenAI tools among IT professionals (97%), with significant perceived gains in personal productivity and organizational efficiency. The findings highlight important correlations between organizational AI adoption and efficiency improvements, while also revealing a concerning relationship between AI investment and employee job insecurity. By identifying key implementation barriers—including output accuracy issues, regulatory compliance challenges, and ethical considerations—this research establishes an empirical foundation for understanding both the transformative potential and inherent challenges of integrating GenAI into software development workflows. Future survey efforts will broaden and diversify participation to produce more robust, generalizable insights. These next steps are essential for informing balanced, evidence-based strategies and frameworks that help IT organizations leverage Generative AI's advantages while safeguarding employee development, work culture and job stability.

**Preprint Notice**